\documentclass[aps,prb,citeautoscript,preprint]{revtex4}

  \usepackage{enumerate}
  \usepackage[pdftex]{graphicx}
  \usepackage{subfigure}
  \usepackage{bm}
  \usepackage{amssymb}
  \usepackage{amsmath}
  \usepackage{eucal}
  \usepackage{float}
  \usepackage{xfrac}
  \usepackage{xcolor}
  \usepackage{fix-cm}
  \mathchardef\mhyphen="2D

\begin{document}

  \title{Directed self-assembly  of   spherical caps via confinement}

  \author{Carlos Avenda\~no}
  \affiliation{School of Chemical and Biomolecular Engineering, 120 Olin Hall, Cornell University,
  Ithaca, New York 14853, USA.}
  \affiliation{School of Chemical Engineering and
  Analytical Science, University of Manchester, Sackville Street, 
  Manchester M13 9PL, UK.}
  \author{Chekesha M. Liddell Watson}
  \affiliation{Department of Materials Science and Engineering, 128 Bard Hall, Cornell University,
  Ithaca, New York 14853, USA.}
  \author{Fernando A. Escobedo}
  \email{fe13@cornell.edu}
  \affiliation{School of Chemical and Biomolecular Engineering, 120 Olin Hall, Cornell University,
  Ithaca, New York 14853, USA.}

  \begin{abstract}
  In this work  we use Monte Carlo simulations to study 
  the phase behavior of spherical caps confined
  between two parallel hard walls separated by a distance $H$.
  The particle model consists of  a hard sphere of diameter $\sigma$ 
  cut off by a plane at a height $\chi$, and it 
  is loosely based on mushroom cap-shaped particles  whose
  phase behavior was recently studied experimentally
  [E. K. Riley and C. M. Liddell, {\it Langmuir}, {\bf 26}, 11648 (2010)].
  The geometry of the particles is characterized by the reduced height
  $\chi^*=\chi/\sigma$, such that the model extrapolates between hard spheres for
  $\chi^*\rightarrow 1$ and infinitely thin hard platelets for $\chi^*\rightarrow 0$.
  Three different particle shapes are investigated: 
  (a) three-quarter height 
  spherical caps ($\chi^*=\sfrac{3}{4}$), 
  (b) one-half height 
  spherical caps or  hemispheres ($\chi^*=\sfrac{1}{2}$), 
  and (c) one-quarter height 
   spherical caps ($\chi^*=\sfrac{1}{4}$). 
  These three models are
  used to rationalize the effect of particle shape, obtained by cutting off spheres
  at different heights,  on the
  entropy-driven  self-assembly of 
  the particles  under strong confinements; {\it i.e.}, for $1<H/\chi<2.5$.  
  As $H$ is varied, a sequence  of crystal structures are 
  observed, including  some having similar symmetry  as that of the structures 
  observed in confined hard spheres on account of
  the remaining spherical surface in the particles, but with additional features
  on account of the particle shapes having intrinsic anisotropy and
  orientational degrees of freedom.
  The $\chi^*=\sfrac{3}{4}$ system is
  found to exhibit a phase diagram that is 
  most similar to the one obtained experimentally for 
  the confined mushroom cap-shaped colloidal particles under.
  A qualitative global phase diagram is constructed that helps reveal the
  interrelations among different phases for all the particle shapes and
  confinements studied.  
 \end{abstract}

  \maketitle

  \section{Introduction}

  Materials that are able to control and modify light propagation are very
  promising for photonic applications \cite{joa08}. 
  Among these, Photonic bandgap (PBG) crystals, which 
  consist of materials having different dielectric constants arranged in a
  periodic structure \cite{xiaam00,hamacie03}, have been the subject of 
  intensive research during the last few decades. 
  PBG crystals can be fabricated by exploiting the self-assembly of 
  polymers, block co-polymers, and colloidal particles  into different 
  morphologies\cite{xiaam01}. In PBG crystals formed by 
  the self-assembly of colloidal particles, the geometry of the  particles
  has been envisioned as one of several possible factors that can modify, 
  and possibly  enhance, the optical properties of the materials
  \cite{luam01,velapl02,glonat07,yanjmc08,hosafm10,hoslan10,foracsnano11,solcocis11}. 
  Clearly, the colloidal particles used for this purpose should have specific
  and well defined shapes and low particle-size polydispersity \cite{funjmc12}. 
  Advances in new routes of synthesis have made possible the fabrication
  of such anisotropic particles in large quantities 
  \cite{lovnl02,kimjacs06,badjacs07,herjpcc07,saccocis11}, so it has become
  possible to produce materials with specific properties by tailoring the
  individual particles. 

  Several examples of crystals formed by the self-assembly of anisotropic particles have been
  reported in the literature.
  Dimers, formed by fusing two lobes of either equal (symmetric) or different (asymmetric)
  diameters, constitute one of the simplest geometries to model anisotropic colloids. 
  PBG crystals formed by symmetric \cite{hoslan10,foracsnano11} and asymmetric 
  dimers \cite{hoslan07,hoslan10} have been reported. 
  The theoretical PGB for this geometry has also been studied \cite{hoslan10},
  showing that the  optical properties of the material can change 
  by modifying the  symmetry and the degree of
  interpenetration of the lobes.
  The influence of particle shape on PBG  has also been 
  studied in a system consisting of spherical silica particles, 
  whose shape can be modified into ellipsoids using ion
  radiation\cite{snoam00,velapl02}. Assuming that the
  material of the particles before and after the radiation is not modified, the PBG
  was found to change as a consequence of the anisotropic shape and the lattice 
  spacing of the new structure.

  Over the years, geometrical confinement of colloidal particles has been used
  as a method to control their self-assembly into 
  well defined crystals structures
  \cite{lowjpcm09}.  For example, in systems formed by
  spherical particles confined between two planar walls or in a wedge, 
  a cascade of different
  phases is observed as a function of the degree of confinement. Experiments
  \cite{pieprl83,nesprl97,ramsm09} and
  computer simulations \cite{schprl96,schpre97,forjpcm06} 
  of spherical particles concur that
  as the confinement is reduced
  (i.e. as the separation between the confining planes is increased) 
  the crystal structures follow the sequence: 
  $1\triangle \rightarrow 1\mathcal{B} \rightarrow 2\Box
  \rightarrow 2\mathcal{R} \rightarrow 2\triangle \rightarrow 2\mathcal{P} 
  \rightarrow  \ldots$, where
  the symbols $\triangle$, $\mathcal{B}$, $\Box$, $\mathcal{R}$, and $\mathcal{P}$ denote
  triangular, buckled, 
  square, rhombic, and prismatic lattices, respectively (see note in
  reference \citenum{note1}). 
  Similar research has been carried out by several groups for anisotropic particles,
  under conditions of strong confinement that approached or enacted
  a behavior in (quasi-) two 
  dimensional geometry. Examples of such experimental and computational
  investigations   include studies of two-dimensional (monolayer) 
  systems for hard dimers \cite{wojprl91,leejmc08,gerprl10,leelan09}, 
  hard spherocylinders \cite{batjcp00}, hard polygons 
  \cite{wojcmst04,schpre05,donpre06,zhaprl09,zhapnas11,zhanc12,zhajacs12,avesm12,atkpre12}, 
  superdisks \cite{jiaprl08}, 
  and studies of systems  confined between parallel walls
  for thin hard-rods \cite{cosjcp03} and  octapod-shaped colloids \cite{qinl12}.

\begin{figure}
  \centering
  \includegraphics[scale=0.4]{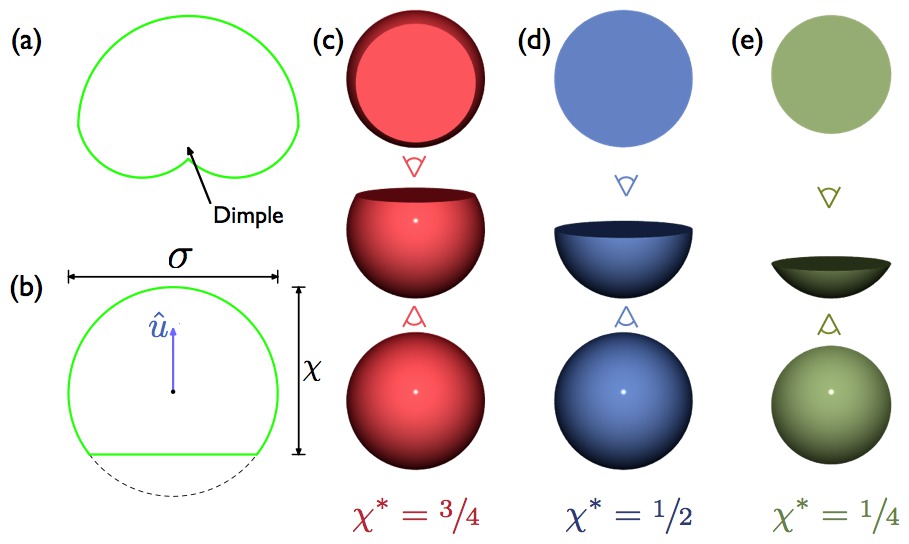}
  \caption{ \label{FigModel} (a) Cross section of an idealized model of
  mushroom cap-shaped particle taken from reference \citenum{rillan10}. 
  (b) Cross section and geometry of the spherical caps 
  particle model studied in this work. 
  The particle model consists of a sphere of diameter $\sigma$ cut off by a plane
  at a height $\chi$. 
  The geometry of the particles is characterized by the reduced height (aspect
  ratio) $\chi^*=\chi/\sigma$, and the orientation is defined by the unit 
  vector $\bf\hat u$ along the principal symmetry axis of the particle.
  Depending on the aspect ratio, three different models are
  defined: (c) three-quarter height spherical caps ($\chi^*=\sfrac{3}{4}$), 
  (d) one-half spherical caps or hemispheres
  ($\chi^*=\sfrac{1}{2}$), and (e) one-quarter height spherical caps
  ($\chi^*=\sfrac{1}{4}$).
  For each model the bottom, side and top views are shown.}
  \end{figure}

  Recently, anisotropic particles 
  whose shape resembles the form of a mushroom cap, have been the object of
  several recent studies
  \cite{hoslan07,rillan10,xumac10,funjmc12,kimjacs12}. 
  The cross section of an idealized model of these particles 
  is show in Figure \ref{FigModel}(a). 
  It has been observed experimentally that 
  under strong confinement such  mushroom 
  cap-shaped particles
  exhibit the formation of the following
  sequence of crystal structures as the 
  confinement is reduced \cite{rillan10}:
  $1\triangle_r \rightarrow 1\mathcal{B}_o \rightarrow 1\mathcal{S}
  \rightarrow 2\Box_o \rightarrow 2\triangle_r \rightarrow 2\mathcal{S}$, where
  $\mathcal S$ denotes a side rotator hexagonal crystal phase, 
  and the subscripts $o$ and
  $r$ refer to the orientations of the particles in each layer; the former
  describes lattices having all the particles oriented in the same direction,
  while the latter denotes random orientations of the particles.
  Although the crystal
  lattices formed by mushroom cap-shaped and spherical particles might appear similar, important
  differences can be observed as a consequence of the orientation of the
  particles in the crystals. In the $1\triangle_r$ and $2\triangle_r$ crystal
  structures, 
  for example, the  particles in each layer are placed in a triangular lattice
  with random orientations that can be either dimple-up or 
  dimple-down, while in the $1{\mathcal B}_o$ crystal phase a bifurcation in the
  particles orientation (BPO) is observed with the formation of alternating 
  stripes of particles oriented dimple-up and dimple-down with either straight 
  or zig-zag buckling. 
  The BPO of the alternating stripes in the $1{\mathcal
  B}_o$ maximizes the 
  packing of the crystal structure \cite{rillan10,riljap12}. As in the case of
  confined hard spheres, the $1{\mathcal B}_o$ transforms into a
  $2\Box_o$ crystal phase showing also a BPO as the confinement is decreased.
  In both $1{\mathcal B}_o$ and $2\Box_o$ the orientation of the particles in each 
  layer is such that the dimples always face towards the walls. 
  In  the $1\mathcal S$ and $2\mathcal S$ 
  rotator crystal phases, the particles 
  in each layer are placed in a triangular lattice
  with the particles rotating in their lattice positions, but keeping
  the  dimples mostly oriented parallel to the walls. 
  As a consequence of the orientation and shape of the particles in the
  $1\mathcal S$ phase, the hexagonal order is weaker than
  that in $\triangle$ phases. 
  Recently, the same group studied the self-assembly of similar mushroom-cap
  shaped particles synthesized using seed emulsion polymerization \cite{funjmc12}. 
  The phase behavior of the new particles was found to be slightly different
  from that previously reported  \cite{rillan10}, in particular,
  in the buckled monolayer $1\mathcal B^*$ 
  the particles were now mainly oriented parallel to the walls, and   in the
  square bilayer $2\Box^*$ the particles in the bottom layer were oriented
  parallel to   the walls while the particles in the upper layer are dimple-up
  oriented. 
  These results suggest that the phases formed by the 
  particles are sensitive to the shape of the
  particles and, possibly, kinetic traps near metastable states.

  Computer simulation can be used to systematically study the effect of
  particle shape, particularly of orientational degrees of freedom, 
  on the self-assembly and phase behavior of anisotropic particles.
  Simulation studies of anisotropic particles having curved
  surface area have mostly focused on bulk behavior.
  For example, Cinnacchi {\it et al.} \cite{cinjpcl10}, reported the bulk phase
  behavior  of contact lens-like particles
  using  Monte Carlo (MC) simulations. 
  For large values of the radius of curvature, approaching the limit of
  infinitely thin hard platelets, the system 
  exhibits the formation of a nematic phase at high densities, 
  consistent with the phase behavior of hard platelets 
  \cite{veepra92}, while for small values of the radius of curvature
  the system shows the formation of clusters with
  spherical-like shape.  The phase behavior of bowl-shaped particles
  (a shape intended to model experimental colloidal particles 
  concurrently studied \cite{marnl10})
  has also been reported based on MC simulations and free energy calculations 
  \cite{marpre10,filprl09}. Simulations of the bowl-shaped model
  reveal the formation of 
  several crystal and liquid crystal structures
  depending on the thickness of the bowls which ranged from thin
  hard-bowls to hard-hemispheres.

  In this work, we present the phase behavior of an idealized model of
  mushroom-cap shaped colloidal particles, which are modelled as 
  spherical caps interacting only via excluded volume, and confined
  between two parallel hard walls. Using this model we address the effect of
  particle shape (as the fraction of curved surface area varies) 
  on the self-assembly and phase behavior 
  for conditions of strong confinement.
  Specifically, our study encompasses
  three different hard-core particle shapes: (a) three-quarter height spherical
  caps,
  (b) one-half spherical caps (hemispheres), and
  (c) one-quarter height spherical caps
  The rest of the paper is organized as follow:
  In section II the simulation model is presented, in section III the
  results for the three different particles shapes, including some general
  remarks are described. In 
  section IV the main conclusion are summarized, followed by an Appendix 
  where a summary of the simulation techniques are described, including the
  algorithm to determine particle-particle and particle-wall overlaps.

  \section{Model}

  The particle model consists of spherical caps (SC),  that is,
  hard spheres of diameter $\sigma$ cut-off by a plane at a height $\chi$.
  Particles only experience  excluded volume interactions and 
  are confined between two parallel hard walls
  separated by a distance $H$. A schematic
  representation of the cross section of the particles is depicted in Figure \ref{FigModel}(b). 
  The particle model is fully characterized by its reduced height (aspect ratio)
  $\chi^*=\chi/\sigma$, and its
  orientation is completely described by the unit vector 
  $\bf \hat u$ along the principal axis of symmetry.
  Note that $\chi^*$ is also equal to the {\it spherical surface fraction} defined
  as the ratio of the curved surface area of the particle to the surface area of
  the complete sphere, while $(1-\chi^*)$ is the ratio of flat to curved surface
  areas.
  In order to address the effect of $\chi^*$ on the phase behavior of the 
  system, three different values of
  the aspect ratio $\chi^*$ are analysed: (a) three-quarter height SC  
  ($\chi^*=\sfrac{3}{4}$),
  (b) one-half height SC or hemispheres ($\chi^*=\sfrac{1}{2}$), and (c) 
  one-quarter height SC ($\chi^*=\sfrac{1}{4}$).
  Throughout this work, structural and thermodynamic properties are reported in 
  reduced units, where the pressure, packing fraction, and separation between
  the walls are given by:  $P^*=P\nu_p/(k_B T)$, $\eta=N\nu_p/V$, 
  and $H^*=H/\chi$, where
  $P$ is the pressure, $k_B$ is Boltzmann's constant, $T$ is the absolute
  temperature, $N$ is the total
  number of particles, $\nu_p=\pi\chi^2(3\sigma/2-\chi)/3$ represents the volume
  of one particle,
  and $V$ is the total  volume of the system. 
  Details of the simulation techniques used to calculate the phase behavior
  of SC are presented in the Appendix.

  \section{Results}

  \subsection{Spherical caps with aspect ratio
  $\boldsymbol\chi^*=\sfrac{3}{4}$}

\begin{figure}
  \includegraphics[scale=0.45]{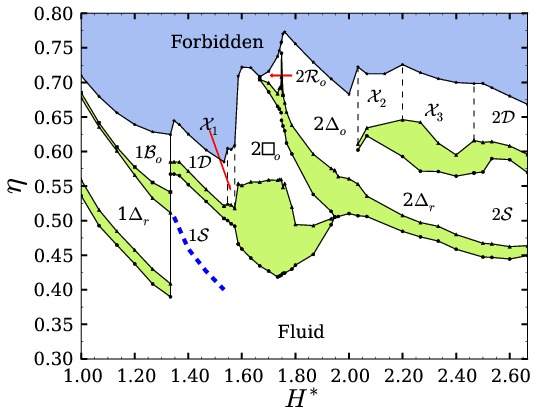}
  \caption{\label{phase075} $\eta\mhyphen H^*$ phase diagram for spherical
  caps with $\chi^*=\sfrac{3}{4}$ obtained from expansion runs using $NPT$ 
  MC simulations. Dashed lines represent approximate phase boundaries.}
  \end{figure}

\begin{figure}
  \includegraphics[scale=0.33]{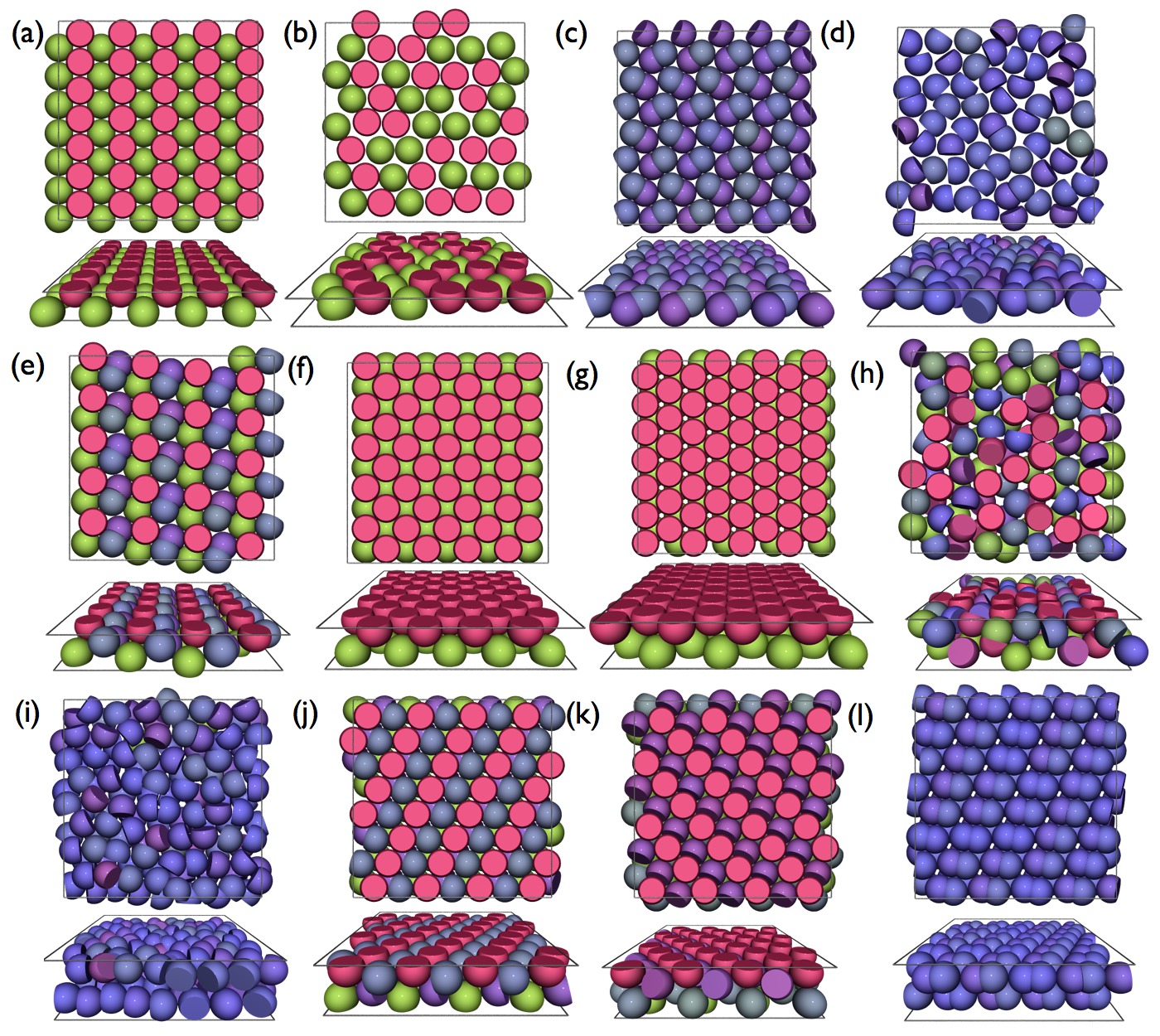}
  \caption{\label{snap075} Representative snapshots for the different phases
  formed by spherical caps
  with $\chi^*=\sfrac{3}{4}$ under confinement. (a) Buckled monolayer
  $1{\mathcal B}_o$, (b) random hexagonal monolayer $1\triangle_r$, (c) dimer
  monolayer $1\mathcal D$, (d) side phase monolayer $1\mathcal S$, (e) 
  intermediate ${\mathcal X}_1$ monolayer, (f) square bilayer $2\Box_o$, 
  (g) hexagonal bilayer $2\triangle_o$, (h) random hexagonal bilayer
  $2\triangle_r$, (i) hexagonal side phase bilayer $2\mathcal S$, (j) and (k)
  correspond to the intermediate $\mathcal X_2$ and $\mathcal X_3$ bilayers, and
  (l) dimer bilayer $2\mathcal D$. The particles are
  coloured according to their orientation with respect to the axis perpendicular
  to the walls. Only a small section of each configuration
  is shown for clarity.} 
  \end{figure}

  The phase diagram for SC with $\chi^*=\sfrac{3}{4}$ under confinement
  is shown in Figure \ref{phase075}. The phase behavior for this system is 
  richer that the one  displayed by confined hard-spheres
  \cite{schprl96,schpre97,forjpcm06}.
  For the highest possible confinements, corresponding to 
  $1.0<H^*\le1.33$ (the upper limit corresponds to $H=\sigma$), the
  particles translate in a quasi-two dimensional fashion adopting only two
  possible orientations: either face-up or face-down.
  The first two stable phases observed at these conditions  are a buckled monolayer 
  $1{\mathcal B}_o$ with BPO and a random orientated hexagonal monolayer $1\triangle_r$.
  Representative configurations are  shown in Figures \ref{snap075}(a-b). 
  As mentioned in the introduction, confined hard spheres also exhibit the
  formation of buckled and hexagonal monolayers, however, the main difference
  is that  SC can adopt several orientations depending of the degree of
  confinement.

\begin{figure}
  \subfigure{\includegraphics[scale=0.38]{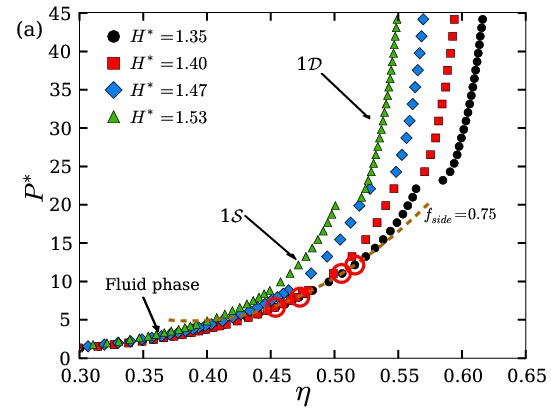}}
  \subfigure{\includegraphics[scale=0.38]{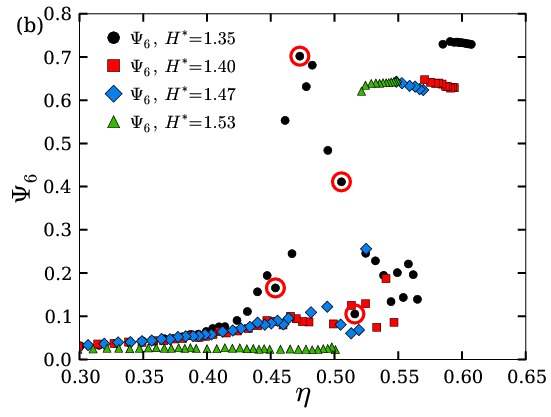}}
  \subfigure{\includegraphics[scale=0.38]{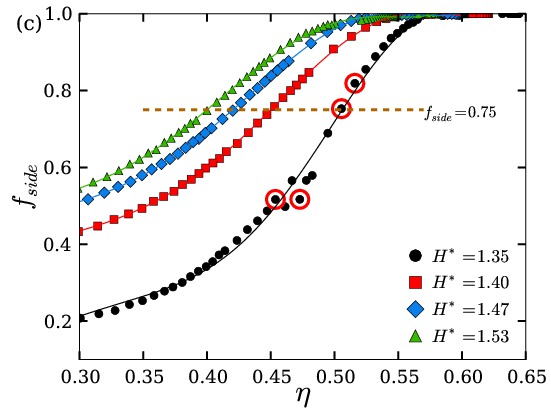}}
  \caption{\label{sideP} (a) Equation of state in the $P^*-\eta$ plane 
  for spherical caps
  with $\chi^*=\sfrac{3}{4}$ under different confinements: $H^*=$1.35, 1.40,
  1.47, and 1.53. 
  (b) Hexagonal order parameter $\Psi_6$ as a function of the packing fraction
  $\eta$ for the same confinements shown in (a).
  (c) Profiles of  the fraction of
  spherical caps oriented parallel to the walls
  $f_{side}$ as a function 
  of $\eta$
  for the same confinements shown in (a) and (b). The lines are polynomial regression
  fits of the simulation points.  The dotted (brown) lines 
  shown in both (a) and (c) denote the approximate boundaries of the 1$\mathcal S$ 
  phase corresponding to states having 75\% of their particles oriented 
  parallel to the wall, i.e., points laying on the constant line of 
  $f_{side}=0.75$  in (c).
  The points in (a), (b) and (c) marked with a large
  (red) open circle correspond to the states shown in Figure \ref{sfactor}.}
  \end{figure}

  For $1.35\le H^*\le 1.53$, high-density
  systems exhibit the formation of a dimer phase $1\mathcal D$, 
  in which the particles orient parallel to the walls 
  pairing up in dimers that can take on several configurations \cite{wojprl91}. 
  In Figure \ref{snap075}(c) one of the dimer crystal phases  with a herringbone 
  configuration is shown. Upon expansion the  $1\mathcal D$ crystal melts into a
  fluid side phase monolayer $1\mathcal S$ in which 
  the particles remain oriented mainly parallel to the walls. 
  For {$H^*=1.35$ (a confinement corresponding to   hard walls separated by a
  distance $1.01\sigma$}), 
  the system exhibits a re-entrant
  hexagonal ordering behavior upon lowering the concentration: the
  hexagonal order first drops (forming a $1\mathcal S$ phase), then climbs up
  and finally drops down again.  While the $1 \mathcal S$ phase occurs for 
  $1.35 \le H^* \le 1.53$, the hexagonal- order re-entrant behavior is only apparent
  for $H^*$ values near 1.35. The phase behavior at these conditions is shown in
  Figure \ref{sideP}, where the equation of state (EoS) and hexagonal order
  parameter   $\Psi_6$ are plotted as a function of the packing fraction
  $\eta$. The only   discontinuity observed in the EoS is for the
  $1{\mathcal D}\rightarrow{1\mathcal S}$ transition. The low-density branch of
  the EoS,    that encompasses   the $1\mathcal S$ phase, decreases
  monotonically showing no evidence of additional phase transitions. However,
  Figure \ref{sideP}(b) reveals that for $H^*=1.35$     a peak in hexagonal order
  occurs that  spans $0.46<\eta<0.51$. For these states, very long runs  were
  carried out to rule out the possibility of artifacts due to   slow relaxation
  and mobility; also, additional compression runs started from an isotropic
  state reproduced the same    re-entrant behavior.

\begin{figure}
  \centering
  \subfigure{\includegraphics[scale=0.34]{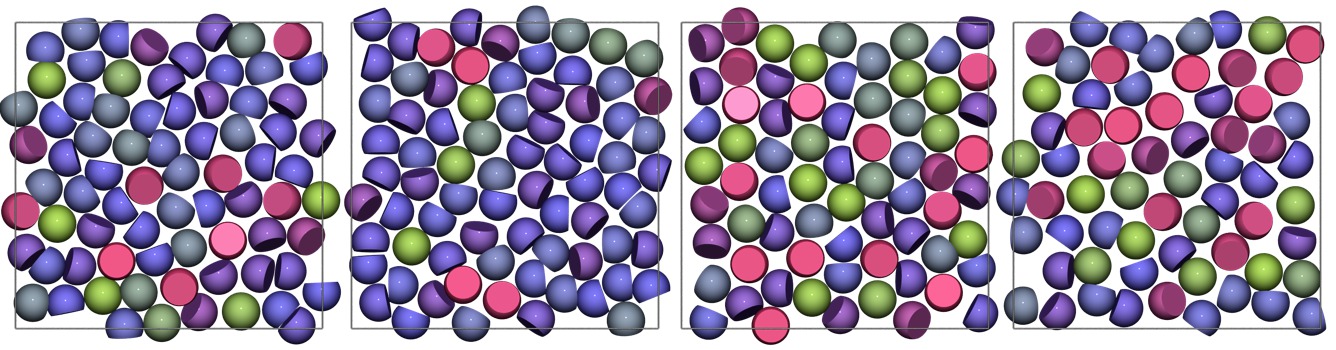}}
  \subfigure{\includegraphics[scale=1.00]{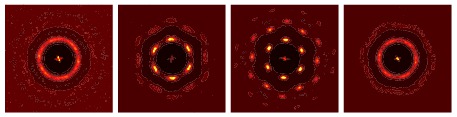}}
  \caption{\label{sfactor} Configuration snapshots and two-dimensional 
  structure factors $S({\bf q})$ for spherical caps
  with $\chi^*=\sfrac{3}{4}$ and 
  $H^*=1.35$. The states correspond
  to the points marked with large (red) open circles in Figure \ref{sideP}.
  From left to right these states are: 
  $P^*=12.1(\eta=0.516,\Psi_6=0.105, f_{side}=0.819)$,
  $P^*=11.1(\eta=0.505,\Psi_6=0.411, f_{side}=0.753)$,
  $P^*= 8.0(\eta=0.473,\Psi_6=0.702, f_{side}=0.517)$, and
  $P^*= 6.6(\eta=0.454,\Psi_6=0.165, f_{side}=0.517)$.
  $S({\bf q})$ is calculated as    
  $S({\bf q})=(1/N)\langle[\sum_{i=1}^N\cos({\bf q}\cdot{\bf r}_i)]^2
  +[\sum_{i=1}^N\sin({\bf q}\cdot{\bf r}_i)]^2\rangle$, 
  where ${\bf   r}_i$   is the position of the centroid of particle
  $i$ and $\bf q$ is a reciprocal wave   vector.
  }
  \end{figure}

  To explore the correlation between side orientation and hexagonal order,
  the fraction of particles
  oriented parallel to the walls   $f_{side}$ is shown in
  Figure \ref{sideP}(c).   A particle $i$ was considered  to   be oriented parallel
  to the wall if $|{\bf \hat u}_i \cdot {\bf \hat u}_w| \leq0.5$,   where ${\bf
  \hat u}_w$ is the unit vector perpendicular to the walls; this angle   ensures
  that the rim of the SC curved surface faces the   walls. It can be observed in
  Figure \ref{sideP}(c) that for $1.35 \le H^* \le 1.53$ the fraction of particles
  side-oriented decreases when the system expands. Figure \ref{sfactor} shows
  snapshots and corresponding two-dimensional structure factors $S({\bf q})$ for
  $H^*=1.35$ in the $0.46<\eta<0.51$ range; it is observed that when most
  particles   remain oriented parallel to the walls (e.g., at $\eta=0.516$)
  the resulting anisotropic shape presented for particle-particle interactions
  frustrates the establishment of hexagonal order,  while when  about half
  particles orient perpendicularly to the walls, they effectively interact as
  spheres which can enhance the   hexagonal order if the volume fraction is
  sufficiently high. It is then a combination of high volume fraction and low
  enough $f_{side}$ that promote hexagonal order. 

   The peak-behavior (and strong
  fluctuations) in $\Psi_6$ values observed in the re-entrant region for
  $H^*=1.35$ can be explained by its proximity to the $1\triangle_r$ phase that
  occurs (in the same $\eta$ range) for $H^*$ just below $H^*=1.33$ (see
  Figure \ref{phase075}); indeed, re-entrant states around $H^*=1.35$ could be
  considered to still be ``part'' of the $1\triangle_r$ phase domain. Of course,
  hexagonal order gets accentuated once $H^*$ is not large enough ($\le 1.33$) to
  accommodate side orientations hence leading to smaller $f_{side}$ values. 
  Re-entrant behavior notwithstanding, it is also difficult to   pinpoint the
  limits of stability of the $1\mathcal S$ phase, but     an approximate
  phase boundary can be sketched  based on the   
  $f_{side}$ vs. density plot (see Figure \ref{sideP}(c)). A threshold of
  $f_{side}=0.75$ is used to draw the   approximate
  phase boundary represented by thick dashed lines in   Figures 
  \ref{phase075} and \ref{sideP}.

  When the confinement is decreased, the $1\mathcal D$ phase transforms into a
  square bilayer $2\Box_o$.  
  This $1{\mathcal D}\rightarrow 2\Box_o$ transformation seems to occurs with a $\mathcal X_1$
  intermediate, which is formed by the separation of some of the dimers in the
  $1\mathcal D$ phase, and their reorientation to face toward the walls (see
  Figure \ref{snap075}(e)). 
  As in the case of confined hard spheres, the
  $2\Box_o$ phase transforms into a $2\triangle_o$ mediated by  a rhombic phase
  $2{\mathcal R}_o$. Likewise $1{\mathcal B}_o$,  these three phases show BPO. 
  As $H^*$ increases at intermediate densities, the  $2\triangle_o$ 
  transforms  continuously into
  a hexagonal bilayer with random particle orientations $2\triangle_r$, which
  then transforms into a hexagonal bilayer $2\mathcal S$ 
  with particles acquiring orientations parallel to the walls. 
  These three phases are shown in
  Figures \ref{snap075}(g-i). This sequence of transformations $2\triangle_o\rightarrow
  2\triangle_r\rightarrow2\mathcal S$ seems to occur continuously as
  no signs of first-order transitions were  observed during the
  simulations.
  
\begin{figure}
  \centering
  \subfigure{\includegraphics[scale=0.38]{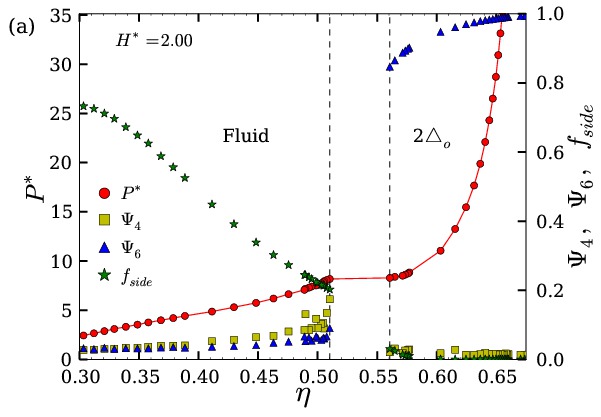}}
  \subfigure{\includegraphics[scale=0.38]{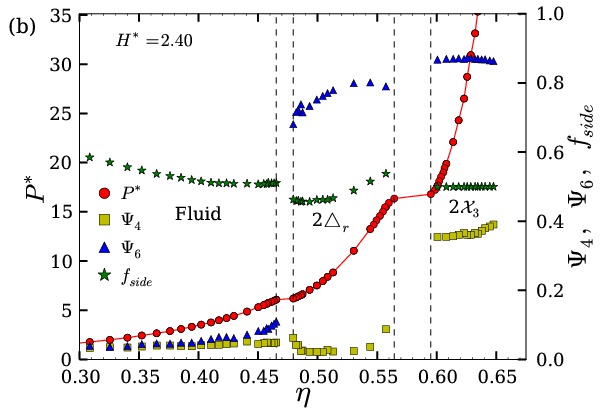}}
  \subfigure{\includegraphics[scale=0.38]{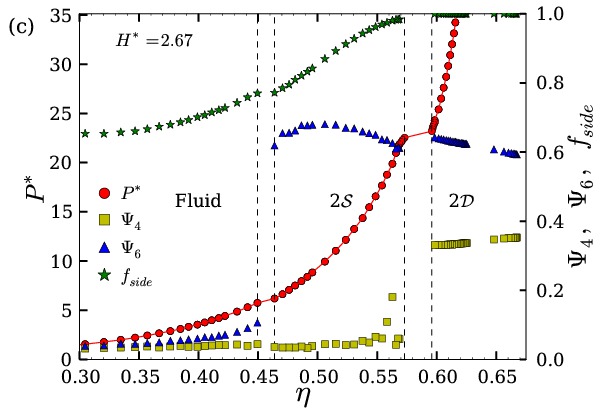}}
  \caption{\label{hex} Equation of state for spherical caps with
  $\chi^*=\sfrac{3}{4}$ for three different confinements: (a) $H^*=2.0$, (b)
  $H^*=2.40$, and (c) $H^*=2.67$. 
  For each confinement the pressure $P^*$, order parameters $\Psi_4$ and
  $\Psi_6$, and fraction of particles oriented parallel to the walls $f_{side}$ 
  are shown as a function of the packing  fraction $\eta$.
  }
  \end{figure}

  At high densities, the $2\triangle_o$ phase transforms
  into a dense dimer side bilayer $2\mathcal D$ structure 
  (shown in Figure \ref{snap075}(l)) as $H^*$ is increased.
  The $2\triangle_o\rightarrow 2{\mathcal D}$
  transformation seems to occur through two intermediaries $\mathcal X_2$ and
  $\mathcal X_3$, which are shown in Figures \ref{snap075}(j-k). In the $\mathcal X_2$
  phase the dimers are
  intercalated between stripes formed by particles with opposite orientations
  facing toward the confining walls. In the second intermediary structure 
  $\mathcal X_3$, the particles forming the dimer in the $\mathcal X_2$ detach
  and reorient to adopt a side configuration, while the
  particles facing toward the confining  walls (in $\mathcal X_2$) remain in a similar
  configuration. Finally, the particles that were facing the walls
  reorient to form the $2\mathcal D$ structure.  
  Upon expansion at constant confinement $H^*$, the $\mathcal X_2$ 
  and $\mathcal X_3$ structures transform into a $2\triangle_r$
  phase, while the $2\mathcal D$ crystal transforms into a $2\mathcal S$ phase.
  These transformation are marked in Figure \ref{hex}, where the equation of state,
  bond order parameters $\Psi_4$ and $\Psi_6$, and fraction of particles
  oriented parallel to the walls 
  $f_{side}$ are shown for three different confinements: 
  $H^*=2.00$, $H^*=2.40$, and $H^*=2.67$. The densest crystal
  structures for each of these confinements are $2\triangle_o$, $2{\mathcal X}_3$,
  and $2\mathcal D$, respectively. 

  The transformation observed upon expansion are: $2\triangle_o\rightarrow {\rm
  Fluid}$ for $H^*=2.00$ (Figure \ref{hex}(a)),  $2{\mathcal X_3}\rightarrow
  2\triangle_r\rightarrow{\rm Fluid}$ for $H^*=2.40$ (Figure \ref{hex}(b)), and
  $2{\mathcal D}\rightarrow 2{\mathcal S}\rightarrow{\rm Fluid}$ for $H^*=2.67$
  (Figure \ref{hex}(c)).
  The three structures $2\triangle_o$, $2\triangle_r$, and 
  $2\mathcal S$ show large values of the bond order parameter $\Psi_6$
  revealing their hexagonal order. However, the $f_{side}$ profiles are different
  for each of the confinements. For the $H^*=2.00$ and $\eta>0.56$,
  $f_{side}$ vanishes indicating that there is a BPO in the structure
  (see Figure \ref{snap075}(g)); however, $f_{side}$ increases once that the fluid
  region has been reached. For the $H^*=2.40$ system the $f_{side}$ profile  shows
  that approximately half of the particles are oriented parallel to the
  walls, 
  confirming that in the hexagonal
  structures $2\triangle_r$ particles have random orientations
  as shown in Figure \ref{snap075}(h). Note that while in the $1\triangle_r$ phase the
  particles can take only face-up or face-down orientations, in the
  $2\triangle_r$ they can adopt any possible orientation.
  Finally, in the $H^*=2.67$ system most of the
  particles are oriented parallel to the walls, 
  as revealed by the large values of $f_{side}$,
  confirming the existence of the $2\mathcal S$ phase. It can also be observed
  that both the $2\triangle_r$ and $2\mathcal S$ phases show moderate values of the
  order parameter $\Psi_6$, while the $2\triangle_o$ phase exhibits larger
  values for this parameter.
  This effect is mainly caused by the particles oriented parallel to the
  wall perturbing the hexagonal order \cite{dulprl06}.

\begin{figure}
  \includegraphics[scale=0.60]{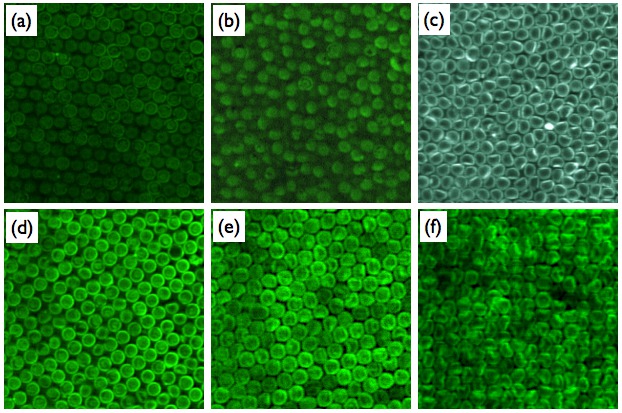}
  \caption{\label{experimental} 
  Experimental mesophases based on the self-assembly of polystyrene ``mushroom
  caps'' under wedge-shaped confinement. (a) $1\triangle_r$, (b) $1\mathcal B^*$, 
  (c) $1\mathcal S$, (d) $2\Box_o$, (e) $2\triangle_r$, and (f) $2\mathcal S$.
  The description of the particle shapes and confocal microscopy
  videos of assemblies are found in references \citenum{funjmc12} and 
  \citenum{rillan10}.
  } 
  \end{figure}

  The SC model with $\chi^*=\sfrac{3}{4}$ presented in this section 
  resembles closely the shape of 
  the mushroom cap-shaped particles reported in
  reference \citenum{rillan10}. The comparison of the structures formed by such 
  SC with those found experimentally reveals that the model is
  appropriate to describe the phase behavior of the real system. As mentioned in the introduction,
  mushroom cap-shaped particles reported in references \citenum{funjmc12} and 
  \citenum{rillan10}
  exhibit the following crystal sequence: 
  $1\triangle_r \rightarrow 1{\mathcal B}_o$ (or $1{\mathcal B}^*$) $ \rightarrow 1{\mathcal S} 
  \rightarrow 2\Box_o $ (or $2\Box^*$) $ \rightarrow 2\triangle_r \rightarrow 2{\mathcal S}$.
  Our model of SC with $\chi^*=\sfrac{3}{4}$ encounters the sequence
  $1\triangle_r \rightarrow 1{\mathcal B}_o \rightarrow 1{\mathcal S} \rightarrow 1{\mathcal D}
  \rightarrow 2\Box_o \rightarrow 2\triangle_r \rightarrow 2{\mathcal S}$ for the horizontal band 
  of packing fractions between $\eta=0.56-0.57$ in Figure \ref{phase075}, suggesting
  that the crystal structures in references \citenum{funjmc12} and \citenum{rillan10} are in
  equilibrium. 
  Figure \ref{experimental} shows the experimental structures for $1\triangle_r$,
  $1\mathcal{B}^*$, $1\mathcal{S}$, $2\Box_o$, $2\triangle_r$, and
  $2\mathcal{S}$ formed by mushroom caps, revealing good agreement with those found in simulation (see Fig. \ref{snap075}) with one exception:
in the experimental buckled monolayer $1\mathcal B^*$ the particles orient with the dimples parallel to the walls, while in the
  simulated $1\mathcal B_o$ the particles orient perpendicular to the
  walls. 
  Moreover, the simulations reveal six additional phases that have not been observed in
  experiments, most of which occur at higher densities and whose structure may not 
  be accessible to experiments due to kinetic trapping. 
  Indeed, the phases observed experimentally appear to be the first ordered 
  structures encountered in our phase diagram (for a given $H^*$) 
  while going up in concentration. It is remarked, however, that the boundaries 
  of the simulated phases are only approximate and  (non-trivial) free energy 
  calculation would be needed to refine them and to determine the stability of
  the additional phases.

  \subsection{Spherical caps with $\boldsymbol\chi^*=\sfrac{1}{2}$}

\begin{figure}
  \includegraphics[scale=0.45]{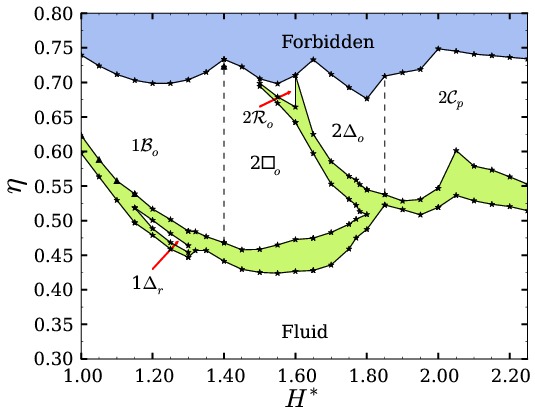}
  \caption{\label{phase050} $\eta\mhyphen H^*$ phase diagram for spherical
  caps with $\chi^*=\sfrac{1}{2}$ (hemispheres)
  obtained from expansion runs using $NPT$  MC simulations. 
  Dashed lines represent approximate phase boundaries.}
  \end{figure}

\begin{figure}
  \includegraphics[scale=0.33]{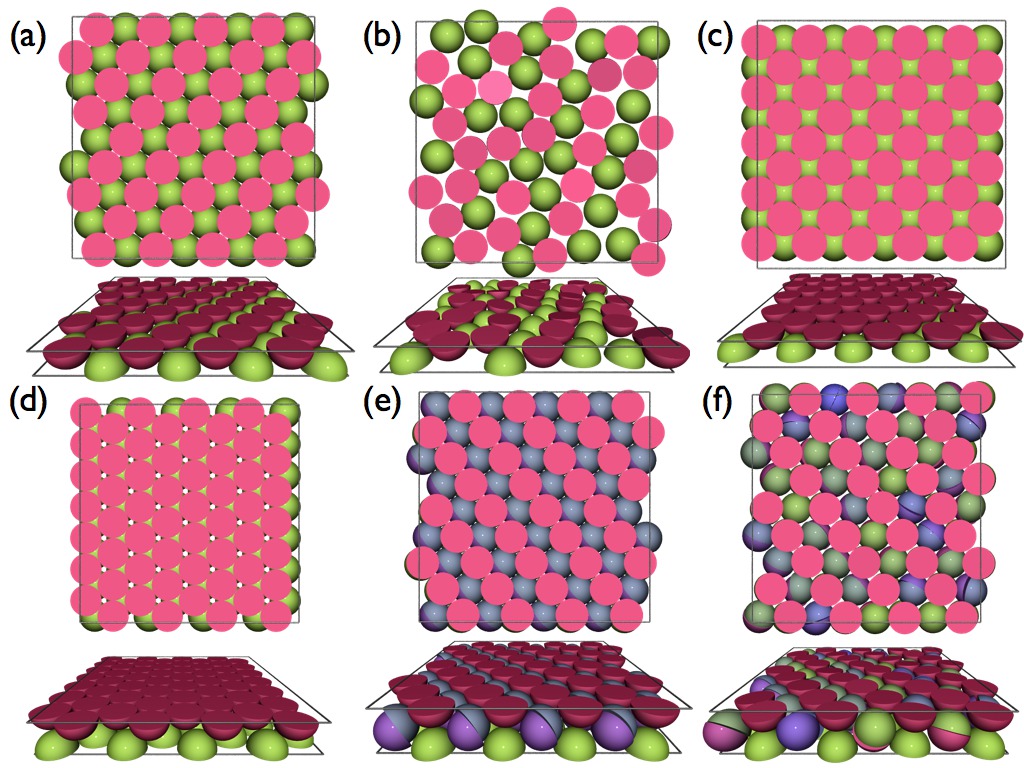}
  \caption{\label{snap050} Representative snapshots for the different phases
  formed by spherical caps with $\chi^*=\sfrac{1}{2}$ (hemispheres) 
  under different confinements. (a) Buckled monolayer
  $1{\mathcal B}_o$, (b) random hexagonal monolayer $1\triangle_r$, 
  (c) square bilayer $2\Box_o$, 
  (d) hexagonal bilayer $2\triangle_o$, and rectangular bilayers 
  $2{\mathcal C}_p$ with dimers in (e) frozen state and (f) plastic rotator
  state.  The particles are
  coloured according to their orientation with respect to the axis perpendicular
  to the walls. Only a small section of each configuration
  is shown for clarity.} 
  \end{figure}

  The phase diagram of confined SC with aspect ratio
  $\chi^*=\sfrac{1}{2}$, corresponding to hard hemispheres, 
  is depicted in Figure \ref{phase050}.
  As in the previous case, the first stable crystal structures observed for this system is
  a $1{\mathcal B}_o$ phase, which is now stable for $1.0<H^*<1.4$. 
  Over a very small region in the phase diagram, it is also observed the 
  formation of a random hexagonal monolayer  $1\triangle_r$, characterized by random
  face-up and face-down orientations of the particles.  
  As in the case of SC with $\chi^*=\sfrac{3}{4}$, this system exhibits the
  formation of three bilayer structures $2\Box_o$, $2{\mathcal R}_o$, and
  $2\triangle$.  In Figure \ref{snap050}
  representative snapshots for these phases are presented.
  For high densities and $H^*> 1.85$, which correspond to wall separations
  larger than the diameter of the particles,  a rectangular
  plastic bilayer phase $2\mathcal C_{p}$ is formed, which is 
  characterized by alternating arrays of
  face-to-face sphere-like dimers,  and column-like dimers. 
  At high densities the orientations of the sphere-like dimers are frozen as
  shown in Figure \ref{snap050}(e); however, when the system is expanded the dimers start to freely
  rotate around their centres of mass as in a plastic rotator phase 
  as  seen in Figure \ref{snap050}(f). 
  The $2{\mathcal C}_p$ phase resembles the ${\mathcal X}_2$ structure seen for
  $\chi^*=\sfrac{3}{4}$, in
  which the dimers are intercalated between stripes formed by particles with
  opposite orientations facing towards the confining wall. However, in the
  ${\mathcal X}_2$ structure the dimers cannot rotate as in the case of
  hemispheres. The 
  ${2\mathcal C}_{p}$ phase also resembles the face-centered cubic rotator phase
  formed by hard-hemispheres at bulk conditions \cite{filprl09,marnl10,marpre10}.

\begin{figure}
  \includegraphics[scale=0.30]{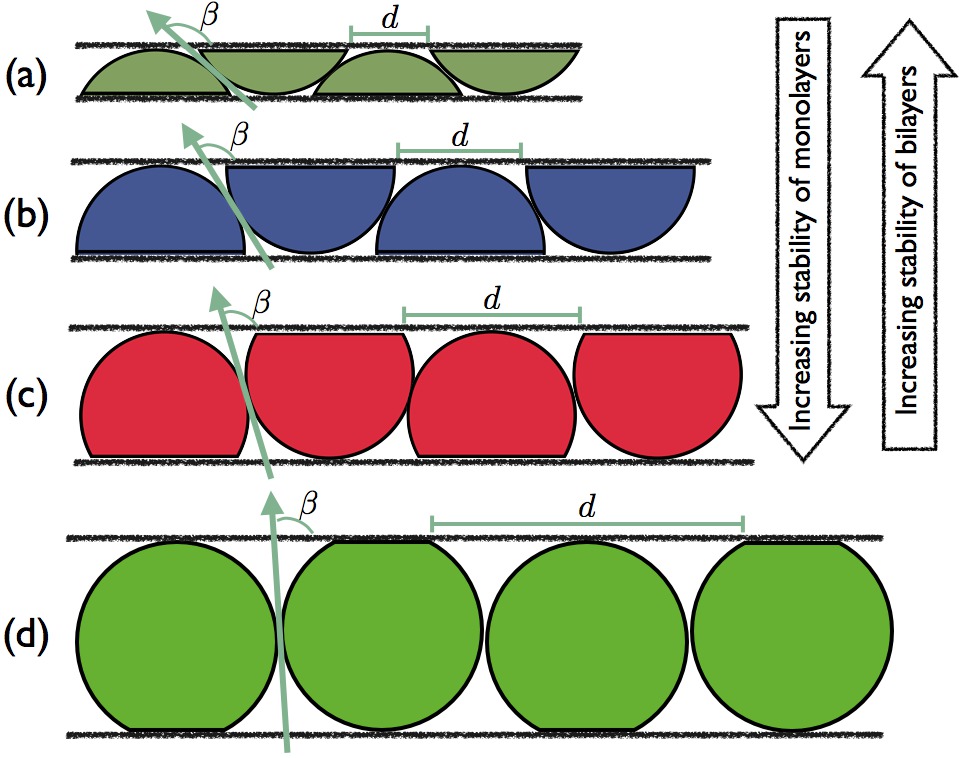}
  \caption{\label{buckledPhase} Buckled crystal structures for maximum
  confinement of
  spherical caps with different aspect ratios: (a) $\chi^*=\sfrac{1}{4}$, (b)
  $\chi^*=\sfrac{1}{2}$,  (c) $\chi^*=\sfrac{3}{4}$, and (d) $\chi^*=\sfrac{1}{95}$. 
  $\beta$ denotes the contact angle between nearest particles in adjacent
  stripes and $d$ is the
  distance between the circumferences of nearest discs in alternating stripes. 
  The formation of an hexagonal monolayer is obtained when 
  $d\rightarrow 2\sigma$ and $\beta\rightarrow90^\circ$ 
  as $\chi^*\rightarrow 1$, which corresponds to the hard-sphere limit.
  Conversely, the formation of a square bilayer is obtained when
  $d\rightarrow 0$ and $\beta\rightarrow 180^\circ$ as $\chi^*\rightarrow 0$, 
  which corresponds to the limit of infinitely thin spherical caps.
  }
  \end{figure}

  The phase behavior of hard hemispheres suggest that the stabilization of
  hexagonal monolayers over large densities and strong confinement 
  is attained only for systems comprised of particles with large
  spherical surface fraction or $\chi^*$.
  This effect can be explained using Figure \ref{buckledPhase} and the following geometrical arguments.
  If $\chi^*$ is increased, the SC bases (capping discs)
  are pushed apart making the monolayers more stable (see
  Figure \ref{buckledPhase}(b)). In this case, the distance $d$ between circumferences
  of nearest discs in alternating stripes increases
  and the contact angle $\beta$ between nearest particles in adjacent
  stripes reduces: the limit values for a perfect monolayer 
  are $d=2\sigma$ and $\beta=90^\circ$. 
  Likewise, it is expected that 
  in SC with $\chi<\sfrac{1}{2}$ under strong confinement, the 
  bases of nearest particles in alternating stripes can come
  closer increasing the stability of bilayers over that of monolayers.
  In particular, it is expected that for  infinitely thin spherical
  caps ($\chi^*\rightarrow 0$) the system would form bilayers directly without
  being preceded by monolayers.
  The same arguments also suggest that the stability of the buckled phase should
  be reduced as $\chi^* \rightarrow 0$.
  These trends are consistent with the results presented  in the following subsection 
  for $\chi^*<\sfrac{1}{2}$ particles.

  \subsection{Spherical caps with $\boldsymbol\chi^*=\sfrac{1}{4}$}

\begin{figure}
  \includegraphics[scale=0.45]{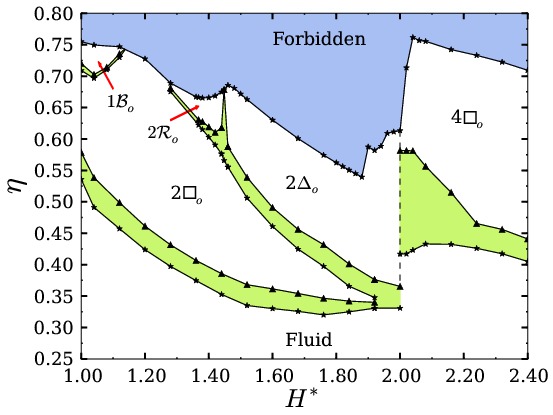}
  \caption{\label{phase025} $\eta\mhyphen H^*$ phase diagram for spherical
  caps with $\chi^*=\sfrac{1}{4}$ obtained from expansion runs using $NPT$
  MC simulations. Dashed lines represent approximate phase boundaries.}
  \end{figure}

\begin{figure}
  \includegraphics[scale=0.33]{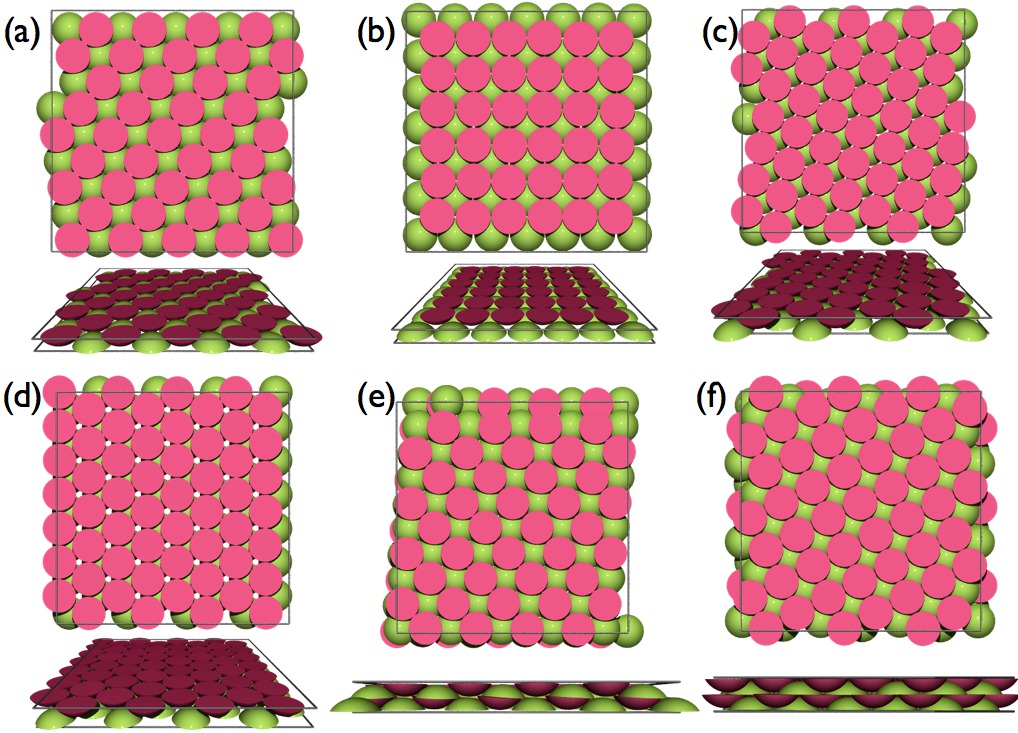}
  \caption{\label{snap025} Representative snapshots for the different phases
  formed by spherical caps with $\chi^*=\sfrac{1}{4}$ 
  under different confinements. (a) Buckled monolayer
  $1{\mathcal B}_o$, (b) square bilayer $2\Box_o$, (c) rhombic bilayer 
  $2{\mathcal R}_o$,
  (d) hexagonal bilayer $2\triangle_o$, and four-layer square phase $4\Box_o$ with
  (e) dimers formation and (f) overlaying alignment. The particles are
  coloured according to their orientation with respect to the axis perpendicular
  to the walls. Only a small section of each configuration
  is shown for clarity.}
  \end{figure}

  The phase diagram for SC particles with $\chi^*=\sfrac{1}{4}$ 
  under different confinements, shown in Figure \ref{phase025}, is simpler than those
  of the other two particle shapes considered in this work.  
  As in the two previous cases, the first crystal phase 
  observed for this system at high densities and strong confinement
  is the buckled monolayer $1{\mathcal B}_o$.
  However, this phase is only observed over a very small region of
  the phase diagram: $H^*<1.20$ and at high densities. 
  This behavior is different from the one observed in confined hard spheres and
  in the previously described SC with $\chi^*=\sfrac{1}{2}$ and
  $\chi^*=\sfrac{3}{4}$, 
  where the buckled phases are stable over a wide range 
  of densities and confinements.
  As a consequence of the small degree of buckling, at high densities the
  $1{\mathcal B}_o$  phase shows either straight or zig-zag stripes \cite{hannat08}.
  In further contrast with hard spheres,  
  SC particles with $\chi=\sfrac{1}{4}$  do not exhibit
  the formation of an hexagonal monolayer, which confirms the destabilization of
  monolayers for small $\chi$ as discussed before in reference to
  Figure \ref{buckledPhase}.
  The small island of $1{\mathcal B}_o$ observed in Figure \ref{phase025} is
  expected to get
  reduced if $\chi^*$ is further decreased. For $H^*<2.0$ the
  most stable crystal phases observed are the $2\Box_o$ and $2\triangle_o$ 
  structures and, as in
  the case of the previous systems, the phase transition $2\Box_o\rightarrow2\triangle_o$ 
  is through a rhombic bilayer $2{\mathcal R}_o$ intermediary. 
  All these bilayer structures, as in the previous cases, also exhibit
  a BPO in such a way that  the base of the SC always face towards the confining
  walls, allowing them to pack efficiently. 
  Representative snapshots for the $1{\mathcal B}_o$,
  $2\Box_o$, $2{\mathcal R}_o$, and $2\triangle_o$ 
  phases are shown in Figure \ref{snap025}. 

  For $H^*\sim 2.0$ the system shows an abrupt change in  phase
  behavior where the number of layers increases from  two to four
  and the structure adopts a square symmetry ($4\Box_o$). 
  In the $4\Box_o$ phase the two middle layers
  can either align to form dimers as in Figure \ref{snap025}(e) or overlay halfway as 
  in Figure \ref{snap025}(f). In this region, it is difficult to 
  pinpoint which crystal structure
  corresponds to a particular confinement unequivocally.
  As in the case of the bilayer structures, the
  $4\Box_o$ if formed by alternating layer of 
  particles with same orientations,
  with the outer layers having only particles oriented facing towards the walls. 

  \subsection{Global remarks on the phase behavior for spherical caps}

\begin{figure}
  \includegraphics[scale=0.50]{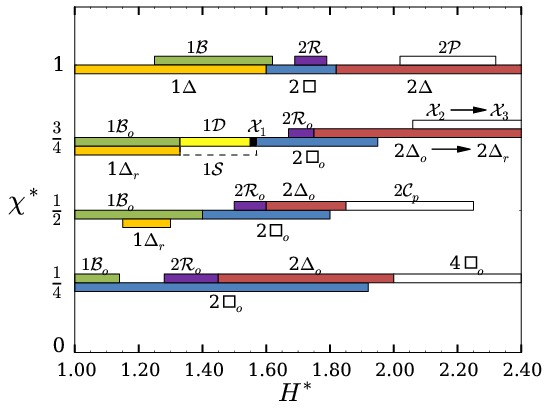}
  \caption{\label{global} 
  Qualitative summary phase diagram of main ordered phases encountered in
  spherical caps with different $\chi^*$ values ($\chi^*=1$ corresponds to hard
  spheres) and different confinements ($H^*$). For any given $\chi^*$, each colored bar
  represents a phase with the bar appearing lower (higher) corresponding to the
  ordered phase occurring at lower (higher) volume fractions.
  The data for $\chi^*=1$ is taken from reference \citenum{forjpcm06}.
  } 
  \end{figure}

  Our results are summarized in the qualitative $\chi^*$ vs. $H^*$ phase diagram presented
  in Figure \ref{global} for the main ordered phases and the $1.0<H^*<2.4$ range only; they are
  presented in the context of the results of the confined hard-sphere system
  ($\chi^*=1$) from reference \citenum{forjpcm06}. Note that hard spheres particles no longer have an
  orientational degree of freedom and hence the phase $1\triangle_r$ simply
  becomes $1\triangle$, $1{\mathcal B}_o$
  becomes $1\mathcal B$ \cite{note1} (denoted as $2\mathcal B$ and I1 in 
  reference \citenum{forjpcm06}), $2{\mathcal R}_o$
  becomes $2\mathcal R$, $2\Box_o$ becomes $2\Box$, and phases
  $2\triangle_o$, $2\triangle_r$, and $2\mathcal S$ fuse into $2\triangle$. 
  The following trends can be observed:

  \begin{itemize}

    \item Two phases that occur in all cases are the bilayers $2\Box_o$ and
    $2\triangle_o$, which are mediated, at very high densities only, by the also
    common phase $2{\mathcal R}_o$. The lower $H^*$ bound in the range where 
    the $2\Box_o$ occurs tends to shift to higher values as $\chi^*$
    increases, while its upper bound remains almost unchanged (at $H^*\sim1.8-1.9$).  

    \item  For $H^* <1.5$, as $\chi^*$ increases the $2\Box_o$ phase gives way to 
    the triangular  monolayer $1\triangle_r$ (or $1\triangle$ for $\chi^*=1$). 
    This process seems to always be mediated at high
    densities by the buckled phase $1{\mathcal B}_o$ (or $1\mathcal B$ for
    $\chi^*=1$). It appears that the $1\mathcal S$ and  $1\mathcal D$ monolayers 
    observed for $\chi^*=\sfrac{3}{4}$ system merge into the $1\triangle$ and
    the $1\mathcal B$, respectively, as $\chi^*$ approaches 1. 

    \item For $2.0<H^*<2.4$, the {$4\Box_o$} phase seen at $\chi^*=\sfrac{1}{4}$ 
    with dimer formation (see Figure \ref{snap025}(e)) appears to be the precursor of 
    the $2{\mathcal C}_p$ phase observed for $\chi^*=\sfrac{1}{2}$, which in
    turn would be the precursor of the ${\mathcal X}_2$ phase observed for
    $\chi^*=\sfrac{3}{4}$, which (together with ${\mathcal X}_3$ and at high 
    densities only) would give way to the $2\mathcal P$
    phase observed for $\chi^*=1$. For intermediate densities, the $2\triangle_r$
    phase observed for  $\chi^*=\sfrac{3}{4}$ likely originates from the
    $2\triangle_o$ phase already occurring for $\chi^*=\sfrac{1}{2}$;
    finally, the $2\triangle_o$ and $2\triangle_r$ seen for
    $\chi^*=\sfrac{3}{4}$ merge into $2\triangle$ phase as $\chi^*\rightarrow 1$. 

  \end{itemize}

  The $2\Box_o$ and $2\triangle_o$ phases are readily 
  rationalized based on considerations of
  ordered packing of particles that (due to confinement) interact laterally with
  their rounded peripheries. The absence of a $1\triangle_r$ monolayer for the
  $\chi^*=\sfrac{1}{4}$ case
  occurs because even for $H^*\rightarrow 1$ the system can accommodate two 
  layers ($2\Box_o$)
  by the alternating flipping of particle orientations. 
  The $1{\mathcal B}_o$ phase can be seen
  as the best packing of particles at high densities when the system confinement
  is such that neither a monolayer nor a bilayer would fit well. Likewise, the
  $4\Box_o$, $2{\mathcal C}_p$, and ${\mathcal X}_2$-${\mathcal X}_3$ 
  phases can be seen as variants of a common motif (where an
  intermediate layer made of ``dimers'' is sandwiched between two monolayers facing
  toward the walls) that occurs as the system tries to pack densely when the
  confinement ( $2.0<H^*<2.4$) is such that neither bilayers nor higher multilayer
  arrangements would work well.

  \section{Conclusions}

  The phase behavior for SC
  under strong confinement inside a slit has been mapped out using MC
  simulations. 
  Depending on the particle aspect ratio and concentration, different
  ordered structures have been observed. The system comprising of SC with
  $\chi^*=\sfrac{1}{4}$  forms stable bilayers with square 
  and triangular symmetry, and  four-layer structures with 
  square symmetry. These systems do not exhibit the formation of stable 
  monolayers, with the exception of  a very small region where a
  buckled monolayer is observed. The stabilization of the monolayers is only
  observed when the spherical surface fraction of the particles $\chi^*$ is increased.
  Indeed, stable buckled monolayers are observed over a wide range of 
  densities for SC with $\chi^*=\sfrac{1}{2}$ (hemispheres) and
  $\chi^*=\sfrac{3}{4}$, 
  while stable random hexagonal monolayers 
  are only  observed for the system with $\chi^*=\sfrac{3}{4}$. 
  A common feature of the three particle
  shapes studied is that for confinements where the particles can 
  fully rotate, the system tends to form dimers to maximize the 
  packing of the structures.
  The progression of phases resulting from changes in confinement and particle
  aspect ratio, can be traced and fully integrated in a global phase diagram
  (Figure \ref{global}), which further provides a seamless connection to the known
  phase behavior of confined hard spheres.  

  The system of SC with $\chi^*=\sfrac{3}{4}$ shows a phase behavior similar 
  to the one found experimentally for  mushroom cap-shaped particles reported in
  reference \citenum{rillan10},
  including the formation of mono and bilayers with particles having
  orientations parallel to the walls (side phases). The
  formation of these structures is only possible when the particles
  have a shape close to spherical.  
  The sequence of structures observed for this system is:
  $1\triangle_r \rightarrow 1{\mathcal B}_o \rightarrow 1{\mathcal S} 
  \rightarrow 2\Box_o \rightarrow \triangle_r \rightarrow 2{\mathcal S}$.
  Our study reveals that such SC can form other structures 
  (at higher densities) that have not been
  observed in experiments yet. Some of the phase boundaries are difficult to pinpoint,
  and future free energy calculations can help elucidate more completely the
  phase diagram for these systems, and the stability of some of the structures. 
  Attempts to obtain 
  the structures by compression runs have met with partial success only,
  revealing the presence of potential kinetic traps. 
  Elucidating the kinetics of some of these order-disorder phase transitions 
  using specialized simulation techniques is part of our ongoing efforts.

  \section*{Appendix}
  \subsection*{Simulation details}

  \renewcommand{\theequation}{A-\arabic{equation}} 
  \setcounter{equation}{0}

  The phase diagrams for each particle shape are obtained using MC
  simulations, and consist of two stages. The first stage corresponds to the
  prediction of the densest crystal structures for each pair of parameter values ($\chi^*,H^*$).
  The second stage consists of expansion runs (in the isobaric-isothermal ensemble) 
  from the crystal structures obtained
  in the previous stage, and analysing the different structures
  observed in the system to assign their phase identity. 

  The calculation of the crystal structures is carried out using the so-called
  floppy-box Monte Carlo algorithm \cite{filprl09,grajcp12}. 
  In this method, systems comprised of a small number of particles  are 
  simulated using  isothermal-isobaric ($NPT$) ensemble MC runs,
  allowing the area and shape of the box 
  in the $x$-$y$ plane to change, keeping the height $H^*$ along the $z$-direction 
  constant \cite{frenkel02}. The methodology entails first
  the generation of a low density configuration, with particles in random positions and
  orientations. The system is then  equilibrated at the pressure of
  $P^*=2.5$, and subsequently compressed in steps of $\Delta P^*=2.5$ until the 
  a value of $P^*=10$ is reached, and subsequent values increased by a factor of 10 
  until  a value of $P^*=10^6$ is reached. In some systems, the compression
  between $P^*=10$ and 100 was carried out using steps of $\Delta P^*=10$, to allow
  more time for the particles to reorganize, which 
  is particularly important for confinements close to the boundaries between
  different crystal phases. The need of a gradual compression was also
  pointed out in reference \citenum{marpre10} for the study of the crystal structures of
  hard-hemispheres at bulk conditions. A set of 50 to 100 of such compressions with
  different initial configurations are performed, and  the last configuration
  in each compression series is recorded for post-processing analysis. 
  Several
  system sizes were tested in order to ensure that the crystal structures are not
  affected by an insufficient number of particles during the calculation,
  however it was observed that using at least 4 particles per layer was enough
  to obtain reliable results: larger systems (more particles per layer)
  show similar structures as those with system of 4 particles per layer.
  In the post-processing analysis several order parameters were used to
  characterize the crystal phases. The symmetry of each layer is obtained
  using bond order parameters, defined as:
  $\Psi_k=\frac{1}{N}\sum_{j=1}^N\frac{1}{n_j}\sum_{k=1}^{n_j}\exp(in\theta_{jk})$, 
  for $n=$4, 6, and, 8, where $\theta_{jk}$ is the angle made by the bond
  between particles $j$ and its nearest neighbor with respect to an arbitrary
  axis, and $n_j$ is the total number of nearest neighbors of particle $j$.
  The calculation of the nearest neighbors is carried out using the Voronoi
  tesselation \cite{Allen87}, except for the case $n=4$, where only the four closest neighbors
  are used. The reason of this is to avoid the well-known problem of degeneracy
  in the Voronoi tessalation in square lattices. The square
  symmetry is easily characterized using the order parameters $\Psi_8$ taking
  into account  all the nearest
  neighbors \cite{mazepl08}; however $\Psi_4$, calculated with the above definition, 
  in combination with $\Psi_6$ are useful to characterize layers with rhombic
  (oblique) and rectangular symmetries. 
  In addition, the following angular order parameters were calculated:
  $S_1=\sum_{i=1}^N\cos(\theta_i)$ and
  $S_2=\sum_{i=1}^N(\frac{3}{2}\cos^2(\theta_i)-\frac{1}{2})$, where $\theta_i$
  is the angle between the orientation of particles ${\bf \hat u}_i$ and the
  main director. The main director of the
  system is obtained from the eigenvector that
  corresponds to the largest eigenvalue of the so-called Saupe tensor defined as: 
  $Q_{\alpha,\beta}=\frac{3}{2}\langle u_{i,\alpha}u_{i,\beta}\rangle - \frac{1}{2}$,
   with $\alpha,\beta=x,y,z$ \cite{lowejp02}. 
   The order parameters $S_1$ and $S_2$ are useful to determine whether or  not a system
   shows a bifurcation in  the particles orientation. For example $S_2$ allows
   to distinguish between $2\triangle_o$ (large values of $S_2$) and
   $2\triangle_r$ (low or moderate values of $S_2$). 
   The structures showing the higher order, measured by the aforementioned order
   parameters, are selected as the crystal structures, which in 
   most of the cases also correspond to the  densest structures.
   However, non-trivial free
   energy calculations may be required to unambiguously determine the most
   stable crystal structures for each confinement, especially to pinpoint
   the boundaries between different phase domains.

  Once the crystal structures are determined, we proceed to perform expansion
  runs to obtain the phase diagram for each particle shape and confinement. As
  in the previous calculation, the expansion runs were carried out using $NPT$ 
  MC simulations. During the simulations involving ordered structures
  the shape of the box is allowed to change, keeping the values of $H^*$
  constant.
  The number of particles $N$ used for each system range between 600 and 700.
  The system is started at high pressures
  and subsequently expanded until the system
  is completely isotropic. One MC cycle is defined as $N$ moves
  consisting of translation moves, rotations and inversions of the particle
  orientations, as well as
  changes of the volume and/or shape of the system box. Each move is randomly selected
  using the following probabilities: 35\% of translation moves, 35\% of
  rotations, 25\% of orientation flips, and 5\% of volume change (50\% of which
  correspond to isotropic changes and 50\% for simultaneous changes 
  of the volume and  shape of the simulation box).
  For each state, $2.5\times10^5$ MC
  cycles are used to equilibrate the system, and additional $1\times10^6$
  cycles, divided in 10 blocks, are used to collect ensemble averages and
  estimate uncertainties. Once the equilibration run for a specific state is
  finished, that configuration is used as a starting point for the next run at
  a different pressure. The location of phase transitions
  are identified based on discontinuities and inflexion 
  points found in the equation of
  state, and in the order parameters previously defined. 

  \subsection*{Overlap algorithm for spherical caps under confinement}

  The heart of the MC simulation program is the algorithm
  to check for the particle-particle and particle-wall overlaps. To
  perform these tests, each particle is modeled as a void SC (bowl)
  capped with a circular disc, thus the
  particle-particle overlap algorithm
  consists of three elementary tests: bowl-bowl \cite{hejpc90,marpre10}, 
  disc-disc \cite{eppmp84}, and  bowl-disc.
  The particle-wall algorithm can also be separated in
  two elementary tests: bowl-wall and disc-wall. 
  Sketches of the different tests for this algorithm 
  are shown in Figure \ref{overlap}.

  The algorithm is started by checking 
  the overlap between the full spherical surfaces of particles $i$ and
  $j$. If the distance  between  the centers 
  of the spheres $|{\bf r}_{ij}|$ 
  is larger than the diameter of the particle
  ($|{\bf r}_{ij}|>\sigma$) then no overlap between the two particles 
  is possible. If $|{\bf r}_{ij}|<\sigma$ then the elementary tests are
  considered.

  \subsubsection{Bowl-Bowl test}

\begin{figure}
  \includegraphics[scale=0.45]{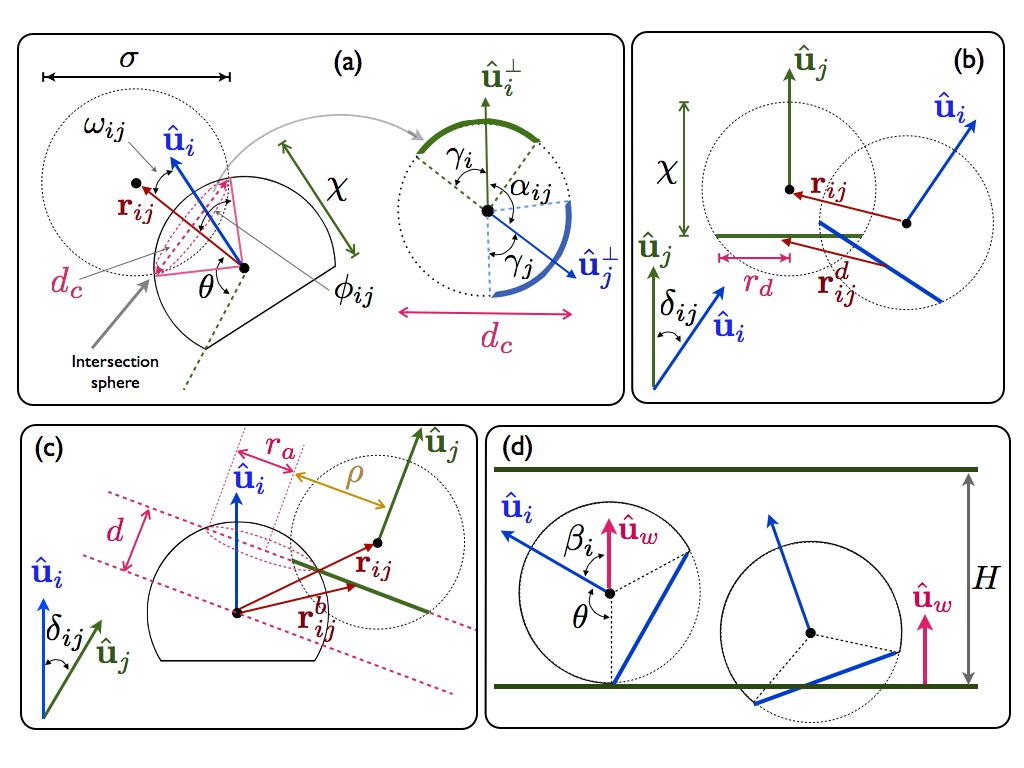}
  \caption{\label{overlap} 
  Description of the overlap algorithm for spherical segments under confinement.
  To implement this algorithm, each particle is modeled as void spherical cap (bowl)
  capped with a circular disc, hence the algorithm comprises 
  the following elementary overlap tests: (a) bowl-bowl \cite{hejpc90,marpre10}, (b)
  disc-disc\cite{eppmp84}, (c) bowl-disc, and (d)  bowl-wall and disc-wall.
  } 
  \end{figure}

  This test is similar to that in reference \citenum{marpre10}, 
  and proceeds as follow: 

  \begin{itemize}
  \item Check first if the bowl of particle $i$ overlaps
  with the sphere of particle $j$. For this purpose 
  calculate the angles $\omega_{ij}$ and $\phi_{ij}$, the former 
  corresponding to the angle between the orientation of bowl $i$, represented by
  the unit vector $\hat{\bf u}_i$, and the interparticle vector ${\bf r}_{ij}$,
  and the latter corresponds to half the opening angle of the cone whose vertex is
  the center of the sphere of particle $i$ and the base is a disc of
  diameter $d_c$ formed by the intersection of the spheres of particles $i$ and $j$.
  The angle $\theta$ (unique for any particle type)
  is also calculated. 
  These three angles, shown in Figure \ref{overlap}(a), are obtained from:

  \begin{equation}\label{bb1}
  \cos(\omega_{ij})=\frac{{\bf r}_{ij}\cdot\hat{\bf u}_i}{|{\bf r}_{ij}|},
  \end{equation}
  \begin{equation}\label{bb2}
  \cos(\phi_{ij})=\frac{|{\bf r}_{ij}|}{\sigma}, \;\; {\rm and}
  \end{equation}
  \begin{equation}\label{bb4}
  \cos(\theta)=\left(\frac{\sigma}{2}-\chi\right)\frac{2}{\sigma}.
  \end{equation}

  The bowl of particle $i$ intersects with the sphere of particle $j$ if

  \begin{equation}\label{bb3}
  |\omega_{ij}+\lambda\phi_{ij}|<\theta,
  \end{equation}

  \noindent for $\lambda=1$ or $-1$. 
  The intersection between the two spheres of particles $i$ and $j$ is a circle of
  diameter $d_c$. Eq. \ref{bb3} holding for both $\lambda$s means that the intersection of
  such a bowl with the sphere of the other particle is also the complete circle of
  diameter $d_c$ (see Figure \ref{overlap}(a)). 
  If Eq. \ref{bb3} holds only once for one $\lambda$ means that
  the intersection is an arc which is a segment of the circle of diameter $d_c$.
  If Eq. \ref{bb3} does not hold for both $\lambda=1$ and $\lambda=-1$ then no overlap is
  possible. Otherwise, if Eq. \ref{bb3} holds for both $\lambda$s and 
  it holds at least once for the bowl of particle
  $j$ (performing a similar analysis), then an overlap exists.
  If Eq. \ref{bb3} holds once for each bowl, then we proceed to the following step.

  \item Both bowls overlap if the intersecting arcs of each bowl
  intersect. Otherwise, the bowls do not overlap. For this,
  angles  $\gamma_i$, $\gamma_j$, and $\alpha_{ij}$ (see
  Figure \ref{overlap}(a)) are calculated from \cite{hejpc90,marpre10}:

  \begin{equation}
  \cos(\gamma_i)=
  \frac{\cos(\theta)-\cos(\phi_{ij})\cos(\omega_{ij})}{\sin(\phi_{ij})\sin(\omega_{ij})},
  \;\;\;{\rm and }
  \end{equation}

  \begin{equation}
  \cos(\alpha_{ij})=\frac{{\bf u}_i^{\scriptscriptstyle \perp} 
  \cdot {\bf u}_j^{\scriptscriptstyle\perp}}
  {|{\bf u}_i^{\scriptscriptstyle\perp}||{\bf u}_j^{\scriptscriptstyle\perp}|}
  \end{equation}

  \noindent where ${\bf u}_i^{\scriptscriptstyle\perp}$ is a vector in the plane
  of the intersection disc that cuts in half the intersection arc of particle
  $i$, and is obtained as:

  \begin{equation}
  {\bf u}_i^{\scriptscriptstyle\perp}=
  \hat{\bf u}_i-\frac{({\bf r}_{ij}\cdot\hat{\bf u}_i){\bf r}_{ij}}{r_{ij}^2}.
  \end{equation}

  The arcs of bowl $i$ and $j$ overlap if

  \begin{equation}
  |\alpha_{ij}|<|\gamma_i|+|\gamma_j|.
  \end{equation}

  \end{itemize}

  \subsubsection{Disc-Disc test}

  For a spherical cap of diameter $\sigma$ and aspect ratio $\chi$, 
  the radius of the capping disc $r_d$ is given by: $r_d=(\sigma^2/4-h^2)^{1/2}$
  (see Figure \ref{overlap}(b)). 
  The distance between the center of the discs of particles $i$ and $j$ is 
  ${\bf r}_{ij}^{\;d}={\bf r}_{ij} + (\sigma/2-\chi)(\hat{\bf u}_j-\hat{\bf u}_i)$.
  Following reference \citenum{eppmp84}, the disc-disc overlap test is
  carried out as follow:

  \begin{itemize}

  \item If the distance between the centers of discs $i$ and $j$
  is larger than the diameter of the discs, $|{\bf r}_{i,j}^{\;d}|>2r_d$, 
  then no overlap is possible (see Figure \ref{overlap}(b)), otherwise the
  test is continued.

  \item Calculate if the plane belonging to the disc $i$ intersects with the plane
  of $j$ and vice versa. If either intersection does not exist, then no overlap
  is possible. This test if performed calculating the distance $a_i$ from 
  disc $i$ to the intersection of the planes of discs $i$ and $j$, and it is
  obtained as

  \begin{equation}
  a_i^2=\frac{({\bf r}_{ij}^{\;d}\cdot\hat{\bf u}_j)^2}
  {1-(\hat{\bf u}_i\cdot\hat{\bf u}_j)^2},
  \end{equation}

  \noindent and likewise for disc $j$:

  \begin{equation}
  a_j^2=\frac{({\bf r}_{ij}^{\;d}\cdot\hat{\bf u}_i)^2}
  {1-(\hat{\bf u}_i\cdot\hat{\bf u}_j)^2}.
  \end{equation}

  Disc $i$ intersects the plane of disc $j$ if $a_i^2<r_d^2$, and so does disc $j$
  if $a_j^2<r_d^2$.

  \item If both discs intersect the planes of each other then we proceed to
  check if the discs overlap. Overlaps between the two discs exists if
  the following inequality holds:

  \begin{eqnarray}
  \left(\frac{|{\bf r}_{ij}^{\;d}|^2}{2}-r_d^2\right)^2
  \sin^2(\delta_{ij})+({\bf r}_{ij}^{\;d}\cdot \hat{\bf u}_i)(\hat{\bf
  u}_i\cdot\hat{\bf u}_j)({\bf r}_{ij}^{\;d}\cdot\hat{\bf u}_j) \\
  < \left[ \left( r_d^2\sin^2(\delta_{ij}) 
  - ({\bf r}_{ij}^{\;d}\cdot \hat{\bf u}_i)^2 \right) 
  \left( r_d^2\sin^2(\delta_{ij}) 
  - ({\bf r}_{ij}^{\;d}\cdot \hat{\bf u}_j)^2 \right)\right]^{1/2} 
  \nonumber
  \end{eqnarray}

  \noindent where $\delta_{ij}$ is the angle between the vectors $\hat{\bf u}_i$
  and $\hat{\bf u}_j$.

  \end{itemize}

  \subsubsection{Bowl-Disc test}

  The bowl-disc overlap test is carried as follow:

  \begin{itemize}
  \item The shortest distance $d$ from the center of the sphere of bowl $i$ 
  to the plane containing the capping disc of particle $j$ is calculated
  (see Figure \ref{overlap}(c)). This distance 
  is given by: $d=|{\bf r}^{\;b}_{ij}\cdot\hat{\bf u}_j|$, where 
  ${\bf r}^{\;b}_{ij}={\bf r}_{ij}+(\sigma/2-\chi) \hat{\bf u}_j$ 
  is the vector from the center of the sphere $i$ to the center
  of disc $j$. If $d>\sigma/2$ then no overlap is possible. If the $d<\sigma/2$
  then we proceed to the following step.
  \item
  The intersection between the plane containing the disc of particle $j$ 
  and the sphere of particle $i$ is a circle of radius
  $r_a=((\sigma/2)^2-d^2)^{1/2}$. We call $\rho$ the shortest distance from the center of
  disc $j$ to the center of the intersection circle of radius $r_a$, which is
  denoted as  $d$ and is calculated as
  $\rho=(|{\bf r}^{\;b}_{i,j}|^2-d^2)^2$. If $(\rho-r_a)>r_d$ then no
  overlap is possible. 
  \end{itemize}

  Note that this test can be optimized by excluding some
  orientations that also include the
  bowl-bowl and disc-disc overlap tests. More specifically, if

  \begin{equation}
  \cos(\theta) < \cos(\delta_{ij})= \hat{\bf u}_i \cdot \hat{\bf u}_j < 1
  \end{equation}

  \noindent then the bowl-disc overlap test can be avoided as this implies that
  disc-disc and/or bowl-bowl can occur along the bowl-disc overlap.

  \subsubsection{Bowl-Wall and Disc-Wall tests}

  Both the bowl-wall and disc-wall overlap can be considered simultaneously in
  the following way:
  \begin{itemize}
  \item We computer the angle $\beta_i$ between the orientation 
  of particle $i$ and the unit
  normal vector perpendicular to the wall $\hat{\bf u}_w$ (see Figure \ref{overlap}(d). 
  This angle is
  obtained as $\cos(\beta_i)=\hat{\bf u}_i\cdot\hat{\bf u}_w$. If
  $\cos(\beta)\le\cos(\theta)$ then the only check that has to be performed is
  the one involving the sphere of
  particle $i$ with the plane of the wall, i.e., if the
  distance $d_w$ from the center of the sphere to the wall is $d_w>\sigma/2$
  then no overlap is possible. 
        
  \item If $\cos(\beta_i)>\cos(\theta)$ and the center of the capping
  disc lays outside the confinement, then an overlap exists.
                             
  \item Check if 
  the disc of particle $i$ and the plane of the wall overlap in analogy
  with with the disc-disc overlap test.
  \end{itemize}

  \section*{Acknowledgements} 
  This work was supported by the U.S. Department of Energy, Office of Basic Energy 
  Sciences, Division of Materials Sciences and Engineering under award Grant No. ER46517.

  \newpage

   \clearpage
   \newpage

\begin{thebibliography}{63}
\bibitem[{not()}]{note1}
\bibinfo{note}{Note that in reference \citenum{forjpcm06} buckled phases are
  denoted as $2\mathcal{B}$, however we adopt the notation $1\mathcal{B}$
  following the convention used in reference \citenum{rillan10} as the buckled
  structures analysed in this work are more conveniently described as
  monolayers.}
\end{thebibliography}

\end{document}